
%
%
\input harvmac

\noblackbox
\Title{\vbox{\baselineskip12pt{\hbox{CTP-TAMU-96/91}}%
{\hbox{hepth@xxx/9205035}}}}
{\vbox{\centerline{Sine-Gordon Solitons as Heterotic Fivebranes}}}

\centerline{HoSeong ~La\footnote{$^*$}{%
e-mail address: hsla@phys.tamu.edu, hsla@tamphys.bitnet}   }

\bigskip\centerline{Center for Theoretical Physics}
\centerline{Texas A\&M University}
\centerline{College Station, TX 77843-4242, USA}
\vskip 1in
The Euclidean analogues of the sine-Gordon solitons are used as sources
of the
heterotic fivebrane solutions in the ten-dimensional heterotic
string theory. Some properties of these soliton solutions are discussed.
These solitons in principle can appear as string-like objects in
4-dimensional space-time after proper compactifications.

\Date{3/92} 

\def\la{\lambda}
\def\half{{\textstyle{1\over 2}}}
\def\d{{\rm d}}
\def\e{{\rm e}}
\def\pa{\partial}

\def\eps{\epsilon}

\font\cmss=cmss10 \font\cmsss=cmss10 scaled 833
\def\IZ{\relax\ifmmode\mathchoice
{\hbox{\cmss Z\kern-.4em Z}}{\hbox{\cmss Z\kern-.4em Z}}
{\lower.9pt\hbox{\cmsss Z\kern-.4em Z}}
{\lower1.2pt\hbox{\cmsss Z\kern-.4em Z}}\else{\cmss Z\kern-.4em Z}\fi}
\def\tr{{\rm tr}}
\def\Tr{{\rm Tr}}
\def\cos{{\rm cos}}
\def\sin{{\rm sin}}
\def\CS{{\cal S}}
\def\CO{{\cal O}}

\vfill\eject
%

The study of solitons has been long pursued in various aspects of physics. It
involves the investigation of nonlinear evolution equations in general.
Now string theory is not an exception anymore.
Lately the structures of the classical solitonic solutions of string theory
have been actively investigated\ref\Fbrev{For recent reviews, see
M.J.~Duff and J.X. Lu, ``A Duality between Strings and Fivebranes,''
Texas A\&M preprint, Class. Quant. Grav. {\bf 9} (1992) 1;
C.G. Callan, J.A.~Harvey and A.~Strominger ``Supersymmetric String Solitons ,''
Chicago preprint, EFI-91-66 (1991); and references therein.}.
In particular, the heterotic fivebrane solution conjectured by Duff\ref\Duf{%
M.J. Duff, Class. Quan. Grav. {\bf 5} (1988) 189\semi
M.J. Duff, in {\it Superworld II}, ed. by A. Zichichi (Plenum, New York, 1990).
}\ and constructed by
Strominger\ref\Stro{A. Strominger, Nucl. Phys. {\bf B343} (1990) 167.}\
exceptionally interesting in the sense that it is dual
to the fundamental string in a generalized sense of the electric-magnetic
duality\foot{This duality which interchanges
Noether charge (e.g. electric charge)
and topological charge (e.g. monopole charge) is in principle the foundation
for
the Montonen-Olive conjecture\ref\MoOl{C. Montonen and D. Olive, Phys. Lett.
{\bf 72B} (1977) 117.}, which is yet to be confirmed rigorously.}.
However, most of the solutions known so far are rather ten-dimensional
solutions
so that their fate in 4-d space-time after compactification is still elusive.

Thus it is important to address a
question that what would be the implications of the
physics of the fivebrane in ten-dimension on the
physics in four-dimensional space-time
after some proper compactification.
There may be some physical consequences
due to the above duality, although such a duality does not necessarily
require the existence of the dual object.
For example, the origin of
the electric-magnetic duality
in four-dimension might be such a string-fivebrane duality in ten-dimension.
In other words,
the monopole solution in four-dimension might be related to the fivebrane
solution in ten-dimension. This aspect was already advocated by Harvey and
Liu\ref\HaLi{J.A. Harvey and J. Liu, Phys. Lett. {\bf 268B} (1991) 40.}.
The dynamical similarities between these two systems are investigated
classically in ref.\ref\Laorb{R. Khuri and H.S. La, Texas A\&M preprint,
CTP-TAMU-95/91, -98/91 (1991).}.
Furthermore, we could conjecture that the solitonic sector in four-dimension
could be
originated from the solitonic sector of ten-dimension.

In this letter, as a first step toward such structures in the
solitonic sector of string theory, we attempt to investigate how these
solutions
appear in the lower dimensional subspaces of
the $(1+9)$-dimensional space-time.
By proper dimensional reductions imposing the Killing symmetries
we make the fields in these subspaces
independent from the rest of the space.\foot{Note that the Killing reduction
does not change the dimensionality of the objects.}

Also in this letter we shall in particular
examine whether sources other than Yang-Mills (YM) instantons can
provide fivebranes. We find that this is indeed possible.
As an example we explicitly work out the case of the
(Euclidean) sine-Gordon solitons. As is well known, the Euclidean sine-Gordon
system does not allow any finite-action static solutions, namely instantons.
But there are infinite-action static solutions, which are good enough for our
purpose.

Solitonic solutions of other systems reduced from the SDYM equation also can be
used to construct fivebranes. To name a few more examples, the hyperbolic
monopoles and vortices are among them. The
detail will be presented elsewhere\ref\Lasolb{H.S. La, ``Solitons reduced
from Heterotic Fivebranes," CTP-TAMU 36/92 (1992).}.
Also we can attempt to analyze the motion
of strings in the fivebrane geometry inside $(1+3)$-dimensional space-time,
using the metric suggested here\ref\Laf{H.S. La, in preparation.}.

\bigskip
\leftline{\bf Heterotic Fivebranes}
\medskip

A fivebrane is a five dimensional extended object and the existence of such
a higher dimensional object is in some sense surprising. Nevertheless, such a
solution exists in string theory. Let us first
review the derivations given in refs.\Stro\ref\CaHaSt{C.G.
Callan, J. Harvey and A.
Strominger, Nucl. Phys. {\bf B359} (1991) 611.}.
The heterotic fivebrane is a solution to the equations of the supersymmetric
vacuum for the heterotic string
\eqn\suei{\delta\psi_M=\left(\nabla_M-{\textstyle {1\over 4}}H_{MAB}\Gamma^{AB}
\right)\eps=0,}
\eqn\sueii{\delta\lambda=\left(\Gamma^A\pa_A\phi+{\textstyle{1\over 6}}H_{AMC}
\Gamma^{ABC}\right)\eps=0,}
\eqn\sueiii{\delta\chi=F_{AB}\Gamma^{AB}\eps=0,}
where $\psi_M,\ \lambda$ and $\chi$ are the gravitino, dilatino and gaugino,
while the anomaly equation is given by
\eqn\anoe{\d H=\alpha' \left(\tr R\wedge R-{\textstyle{1\over 30}}\Tr
	F\wedge F\right)+\CO(\alpha'^2).}
In the above we have properly rescaled all the field variables and that the
string coupling $g_s=\e^{-\phi}$ and $\alpha'$ are the only independent
couplings. In the heterotic string theory $\alpha'$ is proportional to
$\kappa^2/g_{{\rm YM}}$, where $\kappa$ is the gravitational coupling
constant.

The corresponding low-energy effective action for the heterotic string
up to the leading order of $\alpha'$ is
\eqn\eef{\CS={1\over \kappa^2}\int\d^{10}x{\sqrt{-g}}\e^{2\phi}
\left(R+4\pa_\mu\phi\pa^\mu\phi-{\textstyle{1\over 3}}H^2-
{\textstyle{1\over 30}}\alpha'\Tr F^2 +\cdots\right),}
where the dots include the fermionic part of the action that are not relevant
for our purpose now.

The basic idea to have fivebranes is due to the observation
 that the transverse space
to these extended objects is a four dimensional space, which is
conformally equivalent to the Euclidean space, and they should appear as
usual particle-like objects in such a transverse space. Thus it is convenient
to work on the decomposed space $M^{1,9}\to M^{1,5}\times M^4$. Note that for
solitons the Poincar\'e symmetry is not necessarily required, which makes such
a
decomposition possible.

In $(1+9)$-dimension we have Majorana-Weyl fermions, which decompose down to
chiral spinors according to SO(1,9)$\supset$SO(1,5)$\otimes$SO(4) for
such a decomposition. For such spinors the dilatino
equation eq.\sueii\ is satisfied by
\eqn\ei{H_{\mu\nu\la}=\pm\eps_{\mu\nu\la\sigma}\pa^\sigma\phi,}
where $\mu,\nu,...$ are indices for the transverse space
$M^4$ and $\phi=\phi(x^\mu)$, while we shall use indices $a, b,...$ for
$M^{1,5}$.

Then other equations are solved by constant chiral spinors $\eps_{\pm}$ and
\eqn\eii{g_{ab}=\eta_{ab},\ \ \ g_{\mu\nu}=\e^{-2\phi}\delta_{\mu\nu}}
such that
\eqn\eiii{\eqalign{\delta\psi_\mu&=\left(\nabla_\mu+{\textstyle{1\over 2}}
\Gamma_{\mu\nu}\pa^\nu\phi\right)\eps_\pm=\pa_\mu\eps_\pm=0,\cr
\delta\psi_a&=\nabla_a\eps_\pm=\pa_a\eps_\pm=0,\cr}}
and
\eqn\eiv{\delta\chi=F^\pm_{\mu\nu}\Gamma^{\mu\nu}\eps_\pm=
-F^\pm_{\mu\nu}\Gamma^{\mu\nu}\eps_\pm=0,}
where eq.\eiv\ is achieved using the  instanton configuration for the
(anti)self-dual YM equation in $M^4$
\eqn\eins{F^\pm_{\mu\nu}=\pm\half\eps_{\mu\nu}^{\ \ \ \rho\sigma}
F_{\rho\sigma}^\pm}
for an
SU(2) subgroup of $E_8\times E_8$ or SO(32).

Solutions of eq.\eins\ are basic ingredients to build fivebrane solutions.
For example, the instanton solutions lead to the Strominger's fivebrane
solutions. In this case
the basic fivebrane solution is called the  ``gauge'' solution.
Note that $\phi=\phi(r^2)$ now,
i.e. no angular dependence, where $r^2=\sum(x^\mu)^2$ so that the transverse
space has a rotational symmetry.
With a finite instanton scale size $\lambda$, from eqs.\anoe\ei\ we obtain
\eqn\eexpf{\e^{-2\phi}=\e^{-2\phi_0}+8\alpha'{(r^2+2\lambda^2)\over
(r^2+\lambda^2)^2} + \CO(\alpha'^2),}
where $\phi_0$ is the value of the dilaton at spatial infinity.
Thus we have a fivebrane living in $M^{1,5}$ which is a point-like
object in $M^4$.

Now we would like to call the reader's attention to the fact
that any solution of eq.\eins\ in principle leads to a fivebrane
solution, as long as the anomaly equation eq.\anoe\ provides a nontrivial
solution for the dilaton. In
particular many lower dimensional solutions to the self-dual YM
equation are known\ref\rWa{R.S. Ward,
``Integrable Systems in Twistor Theory,"
in {\it Twistors in Mathematics and Physics}, (Cambridge Univ. Press,
1990).}\ so that in principle we can relate all these solitonic solutions to
the heterotic fivebranes.

\bigskip
\leftline{\bf Euclidean Sine-Gordon Case}
\medskip

\def\zb{{\overline z}}
\def\wb{{\overline w}}
The (anti)self-dual YM equations have an interesting reduction to the
two-dimensional solitonic system, namely the sine-Gordon equation.
Here we shall attempt a new reduction of the (A)SDYM equation to the
Euclidean sine-Gordon equation for the gauge group SU(2) and the
Euclidean signature, then to solve the anomaly
equation eq.\anoe\ for this solution.
For the Euclidean signature we can introduce two sets of complex coordinates
for convenience, although one can use the real coordinates, as
\eqn\comcor{z=x+iy,\ \zb=x-iy,\ \ w=u+iv,\ \wb=u-iv,}
where $(x,y,u,v)$ are the cartesian coordinates.
In this coordinate system the SDYM equations will be written as
\eqn\sdymea{F_{z\zb}-F_{w\wb}=0,\ \ F_{z\wb}=0,\ \ F_{\zb w}=0,}
while the ASDYM equations are
\eqn\asdymea{F_{z\zb}+F_{w\wb}=0,\ \ F_{zw}=0,\ \ F_{\zb \wb}=0.}

For the gauge group SU(2) with the  generators
$J_\pm={\textstyle{1\over {\sqrt 2}}}(J_1\pm iJ_2),\ \ J_3$, which are in the
adjoint representation such that $(J_a)_{bc}=-i\epsilon_{abc}$,
we can introduce an ansatz for the gauge fields as
\eqn\idgau{A_z=f_1 J_3,\ A_\zb=f_2 J_3,\ A_w=g_1 J_+ +g_3 J_-,
\ A_\wb=g_2 J_- + g_4 J_+.}
With such identifications the SDYM equations reduce to
\eqn\nwrds{\eqalign{f_1&=\pa_z\ln g_2=-\pa_z\ln g_4,\cr
			f_2&=-\pa_\zb\ln g_1=\pa_\zb\ln g_3,\cr
		0&=\pa_z f_2-\pa_\zb f_1 -g_1 g_2 +g_3 g_4,\cr}}
and the conditions that $\pa_\wb f_1=\pa_w f_2=0,
\pa_w g_2=\pa_\wb g_3, \ \pa_w g_4=\pa_\wb g_1$. The last conditions can be
simply satisfied by requiring two Killing symmetries along $(u,v)$ directions
that none of the fields depend on the
$(u,v)$-coordinates. For the ASDYM equation we obtain more or less the same set
of equations.

Now defining
\eqn\ancon{g_1=-g_2=e^{-{i\over 2}\psi},\ g_3=-g_4=e^{{i\over 2}\psi},}
we obtain the Euclidean version of the sine-Gordon equation,
\eqn\esigo{\pa_z\pa_\zb\psi-2\sin\psi={\textstyle{1\over 4}}(\pa_x^2 +
\pa_y^2)\psi - 2\sin \psi=0.}
The above is related, redefining  $y=it$, to the $(m^2=8)$
sine-Gordon equation
\eqn\msge{(\pa_t^2-\pa_x^2)\varphi+{m^2\over\lambda}\sin
\lambda\varphi=0,}
where  the coupling constant  $\lambda$ can be
rescaled away since we are not interested in quantizing this system  here.

In this background the anomaly equation eq.\anoe\ becomes up the first order
of $\alpha'$
\eqn\sganoe{(\pa_x^2+\pa_y^2) \e^{-2\phi}=2\alpha'\left[\sin\psi\left(
\pa_x^2+\pa_y^2\right)\psi+\cos\psi\left((\pa_x\psi)^2+(\pa_y\psi)^2\right)
\right].}
Using the above sine-Gordon equation we can easily solve this equation to
obtain a solution
\eqn\bisgol{\e^{-2\phi}=\e^{-2\phi_0}+2\alpha'(1-\cos\psi),}
where $\psi$ satisfies the sine-Gordon equation
and $\phi_0$ is the value of the dilaton $\phi$ at
$x, y=\pm\infty$.

Due to the Derrick's theorem\ref\Derr{G.H. Derrick, J. Math. Phys. {\bf 5}
(1964) 1252\semi R. Hobart, Proc. Phys. Soc. {\bf 82} (1963) 201.}\
applied to the Euclidean sine-Gordon theory, there
is no finite-action static solution for $\psi$. Nevertheless, we can have
infinite-action static solutions, which do not generate any tunnelling effect.
In fact we can easily find the following solution:
\eqn\sgsol{\psi=4Q\tan^{-1}\left[\gamma\e^{\alpha x+\beta y}\right],}
where $\gamma$ is an arbitrary irrelevant constant so that we can set
$\gamma=1$
without loss of generality, and ${\alpha^2+\beta^2}=8$.
$Q=\pm 1$ is the soliton charge.
This solution is related to the soliton solutions of
the sine-Gordon equation eq.\msge,
\eqn\sgkink{\varphi=4Q\tan^{-1}\left[\exp\,
m{x-ct\over\sqrt{1-c^2}}\right],}
identifying
\eqn\resg{y=it,\ \ c=i\tilde{c},\ \ \alpha={m\over\sqrt{1+{\tilde c}^2}},\ \
\beta={m\tilde{c}\over\sqrt{1+{\tilde c}^2}},\ \ m=2\sqrt 2.}

It is straightforward to
show that the corresponding action of the Euclidean sine-Gordon
theory is indeed infinite for these solutions. However, this cannot be a
reason to abandon these solutions for our purpose because this action is not an
essential ingredient for fivebrane solutions.
 Note that the SDYM equation is not an
equation of motion so that the action for any reduced system from the SDYM
equation is not relevant to us.
Due to the self-dual YM
structure, the corresponding fivebrane solutions can still
saturate the Bogomol'nyi bound of the action of the heterotic string theory.
Strictly speaking, the fivebrane is not an instanton related to the tunnelling
effect because we work on the $(1+9)$ dimensional spacetime.
{}From this point of view, whether the action of the heterotic string is finite
or not is not really a relevant issue to us. We are just interested in looking
for some solitonic solutions.

Using eq.\sgsol,  now the dilaton eq.\bisgol\ becomes
\eqn\bslo{\e^{-2\phi}=\e^{-2\phi_0} + 16\alpha' {\e^{2(\alpha x+\beta y)}\over
\left(\e^{2(\alpha x+\beta y)}+1\right)^2}.}
Note that this solution does not have any singularity and depends on the
$x,y$-coordinates explicitly, not just on $x^2+y^2$.
This dilaton solution does not care about
the sign of the soliton charge $Q=\pm 1$, while the YM fields depend on the
charge $Q=\pm 1$.
We can also express the YM fields eq.\idgau\ in terms of eq.\sgsol\ as follows:
\eqn\ymsol{\eqalign{A_z&=-Q(\beta+i\alpha)
{\e^{(\alpha x+\beta y)}\over \e^{2(\alpha x+\beta y)}+1 } J_3,\cr
A_\zb&=Q(\beta-i\alpha)
{\e^{(\alpha x+\beta y)}\over \e^{2(\alpha x+\beta y)}+1 } J_3,\cr
A_w&={1-\e^{2(\alpha x+\beta y)}-i2Q\e^{(\alpha x+\beta y)}\over
	\e^{2(\alpha x+\beta y)}+1}J_+ +
{1-\e^{2(\alpha x+\beta y)}+i2Q\e^{(\alpha x+\beta y)}\over
	\e^{2(\alpha x+\beta y)}+1}J_- ,\cr
A_\wb&={1-\e^{2(\alpha x+\beta y)}-i2Q\e^{(\alpha x+\beta y)}\over
	\e^{2(\alpha x+\beta y)}+1}J_- +
{1-\e^{2(\alpha x+\beta y)}+i2Q\e^{(\alpha x+\beta y)}\over
	\e^{2(\alpha x+\beta y)}+1}J_+ .\cr}}
The fact that there are all the four dimensional YM fields indicates that the
solutions we have here are still  fivebrane solutions.

Now let us count the zero modes.
In the two-dimension parametrized by $(x, y)$ coordinates the soliton solutions
eq.\sgsol\ generate four zero modes, which are two for the two translational
symmetries of the $x, y$-directions, one for the $(\alpha^2+\beta^2=8)$
``scaling" symmetry and one for the O(2) rotational symmetry of $(\alpha
x+\beta
y)$. This last O(2) symmetry is due to the fact that the O(2) rotation of
$(x,y)$ can be compensated by O(2) rotation of $(\alpha, \beta)$.
Since the two Killing symmetries, $(\pa_u,\ \pa_v)$,
generate four extra zero modes for the
fivebrane, the fivebrane solution still has 120 bosonic zero modes, including
112 zero modes due to $E_8\to SU(2)\times E_7$, like in the ``gauge" solution
case. We expect that the fermionic zero modes counting is also similar to the
``gauge" solution case.

\bigskip
\leftline{\bf Discussion}
\medskip

In general one could expect that the fivebranes could appear as particles,
strings or membranes in the $(1+3)$ dimensional spacetime.
In this letter we have presented an explicit construction of solitonic
solutions
in the two dimensional subspace of the transverse space $M^4$,
reduced from the heterotic fivebrane solutions in ten-dimension.
The fivebrane solution we have derived looks like a string in the
$(1+3)$ dimensional space defined by $(x, y)$ plane and a $(1+1)$ dimensional
subspace of $M^{1,5}$. Whether this solution should behave like a cosmic string
is another issue because we have not mentioned anything about the symmetry
breaking structure yet. But after we find out the detail of the
compactification
scheme, this should be an important issue to be addressed.

We also expect that
the origin of the electric-magnetic duality in four-dimensional
world is originated from the string-fivebrane duality in ten-dimension in such
a way that the solitonic sector of the four-dimensional effective field theory
might be coming from the fivebrane sector of the string theory.

{}From this point of view, further study of the properties of these solutions
presented here should be important in future.

\bigbreak\bigskip\bigskip\centerline{{\bf Acknowledgements}}\nobreak

\par\vskip.3truein

The author would like to thank M. Duff, D. Nanopoulos and E. Sezgin
for conversations on related subjects.
This work was supported in part by NSF grant PHY89-07887 and a World Laboratory
Fellowship.


%
\listrefs
\vfill\eject
\bye